\documentclass{vgtc}                          
\graphicspath{{figures/}{pictures/}{images/}{./}} 

\usepackage{times}                     
\usepackage{float}
\usepackage{amsmath}
\usepackage{tabu}                      
\usepackage{booktabs}                  
\usepackage{lipsum}                    
\usepackage{mwe}                       
\usepackage{mathptmx}  
\usepackage{amsmath}

\vgtcinsertpkg

\usepackage{algpseudocode}
\usepackage{mathptmx}                  
\usepackage{bm}                       
\usepackage{amsmath}                  
\usepackage{textcomp}                 

\title{SpaceVLA: Spatially Grounded VLA for Robotic Manipulation with User-Authored Grasp and Place Anchors}

\newcommand{\equalcontribsymbol}{\ensuremath{{}^{a}}}

\author{
Daniia Zinniatullina$^{1,2}$\equalcontribsymbol
\thanks{e-mail: Daniia.Zinniatullina@skoltech.ru}
\and Iaroslav Kolomiets$^{1,2}$\equalcontribsymbol
\thanks{e-mail: Iaroslav.Kolomiets@skoltech.ru}
\and Mikhail Konenkov$^{1,2}$
\thanks{e-mail: Mikhail.Konenkov@skoltech.ru}
\and Miguel Altamirano Cabrera$^{1,2}$
\thanks{e-mail: M.Altamirano@skoltech.ru}
\and Dzmitry Tsetserukou$^{1}$
\thanks{e-mail: D.Tsetserukou@skoltech.ru \\
    \equalcontribsymbol These authors contributed equally to this work.}
}

\affiliation{
\scriptsize
$^{1}$Skolkovo Institute of Science and Technology, Russian Federation.\\
$^{2}$R\&D Center, MWS, Russian Federation.\\
}
\teaser{
  \centering
  \includegraphics[width=1\linewidth]{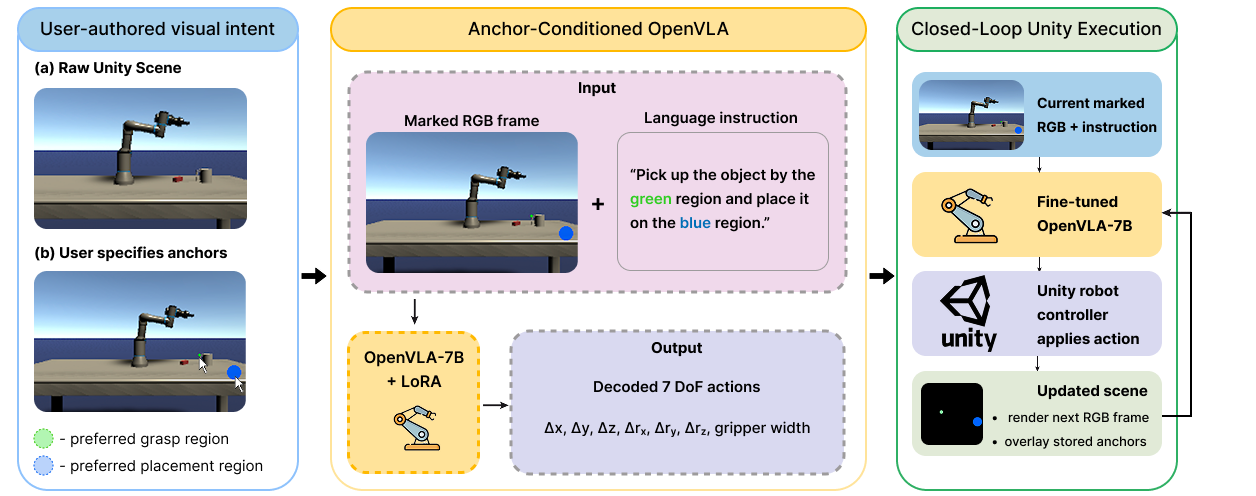}
  
  \caption{Overview of Visual Intent Anchors. A user specifies grasp and placement targets in the Unity interface. The targets are converted into color-coded image-space anchors and overlaid on the RGB observation provided to VLA. The fine-tuned policy predicts 7-DoF incremental actions, which are executed in closed loop in the Unity robot environment.}
  \label{fig:teaser}
}

\abstract{
Vision-language-action (VLA) models follow language commands but often lack explicit spatial intent for manipulation. We present Visual Intent Anchors, an XR pipeline that lets users specify grasp and placement regions and renders them as image-space overlays for VLA control. We collect 200 Unity pick-and-place demonstrations and fine-tune OpenVLA-7B with LoRA on temporally subsampled annotated observations. The policy predicts tokenized 7-DoF incremental actions from marked RGB observations and language. We evaluate the policy in closed-loop Unity trials, achieving a grasp success rate of 91.25\% and mean grasp and placement errors of 0.5 cm and 0.7 cm, respectively.

} 

\keywords{Mixed Reality, Human-Robot Interaction, Vision-Language-Action Models, Robot Manipulation, Visual Grounding, Spatial Interaction.}

\begin{document}


\firstsection{Introduction}

\maketitle

Vision-language-action (VLA) models map visual observations and language instructions to robot actions {\cite{{brohan2023rt2visionlanguageactionmodelstransfer},{kim2024openvlaopensourcevisionlanguageactionmodel}}, but language alone often underspecifies the spatial intent needed for manipulation. For example, an instruction such as ``pick up the mug and place it on the table'' does not specify which part of the mug should be grasped or which location on the table should be used. Multiple grasps and placement locations may be physically feasible, while only one reflects the user's intended interaction.

Existing grounding methods infer feasible interaction regions from language and visual context but may not capture a user’s preferred strategy when several options are valid {\cite{tang2025affordgraspincontextaffordancereasoning}}. XR can provide such spatial input, yet it is typically used for teleoperation, trajectory specification, or demonstration {\cite{zhang2018deepimitationlearningcomplex}}.

In mixed-reality robot programming, users often need to specify spatial preferences without manually teleoperating the full robot trajectory. We introduce \textit{Visual Intent Anchors}, a Unity-based XR interface for preference-aware VLA manipulation. At the beginning of an episode, the user manually annotates a grasp target point on the object and a placement target point in the workspace. These points are converted into stored visual masks and rendered as color-coded image-space overlays in the RGB observation consumed by the policy: green denotes the grasp target and blue denotes the placement target.

We instantiate the framework in Unity, collect 200 pick-and-place demonstrations, and fine-tune OpenVLA-7B with LoRA on 120 training episodes. The user annotates the target points once per episode. After each predicted robot action, Unity renders a new workspace frame and re-applies the stored active masks before the next policy query. The policy receives the updated marked RGB observation and language instruction, predicts a 7-DoF incremental action, and executes it through the Unity robot controller in closed loop.

We evaluate the framework on 80 held-out Unity episodes under three conditions: (1) the anchor-conditioned policy with \textit{intended grasp and placement anchors}, (2) the same policy with \textit{randomly positioned anchors on the objects}, and (3) a \textit{no-anchor baseline} trained and evaluated without visual anchors. With the intended anchors, the policy completes 73 of 80 tasks successfully, achieving a full task success rate of 91.25\%. The mean grasp and placement end-effector position errors are 0.5 cm and 0.7 cm, respectively. When random anchors are provided to the same policy, the success rate decreases to 62 of 80 tasks, or 77.5\%, while the mean grasp and placement errors increase to 2.6 cm and 3.4 cm. The no-anchor baseline achieves 40 of 80 successful episodes (50\%).

We additionally test the policy with anchors placed at arbitrary workspace locations. The robot redirects its motion toward these anchors, showing that the policy uses their spatial position rather than only their visual appearance.

The contributions of this work are:
\begin{itemize}
\item \textbf{User-authored spatial intent interface.}
We introduce a Unity-based spatial authoring interface that allows users to specify preferred grasp and placement targets and communicates them to a VLA policy through visual anchors. In the current prototype, the anchors are represented as fixed image-space masks under a static camera.

\item \textbf{Anchor-conditioned VLA pipeline.} We present a pipeline that converts user-defined target points into temporally rendered visual anchors for training and closed-loop execution of an image-conditioned VLA policy.

\item \textbf{Closed-loop evaluation of spatial conditioning.} We compare correct anchors, randomly positioned anchors on the objects, and a no-anchor baseline, and additionally test arbitrary anchor locations, including positions outside the objects, to verify whether the policy follows the specified spatial targets.

\end{itemize}

\section{Related Work}

\subsection{XR Grounding for Manipulation}

Extended reality interfaces support robot interaction through teleoperation, hand tracking, controllers, and virtual scene representations. Recent systems also use XR for robot learning from demonstration through hand motions or VR controllers {\cite{{iyer2024openteachversatileteleoperation},{cheng2024opentelevisionteleoperationimmersiveactive}}.

Visual grounding and affordance-based manipulation methods identify task-relevant objects, interaction regions, and feasible grasps from visual observations and language instructions {\cite{cheng2024opentelevisionteleoperationimmersiveactive}}. Some approaches use segmentation masks or grounded visual representations to improve target selection and placement. For example, RoboGround {\cite{huang2025robogroundroboticmanipulationgrounded} generates object and placement masks with a grounded vision-language model and uses them for manipulation.

However, these methods typically infer grounding automatically and may not reflect the user’s preferred strategy when multiple valid grasp or placement options exist. Our approach treats grounding as user-authored spatial intent by allowing users to explicitly select grasp and placement regions in a Unity-based spatial authoring interface. This interaction can subsequently be extended to immersive VR and AR interfaces.
\subsection{VLA Models and Conditioning}

Vision-language-action models map visual observations and language instructions to robot actions, enabling generalist manipulation policies trained on large-scale robotic and vision-language data. OpenVLA provides an open 7B-parameter VLA model and supports efficient adaptation to new manipulation tasks through fine-tuning.

Beyond generalist VLA backbones, task-specific systems have demonstrated language-guided bimanual manipulation. Bi-VLA performs dexterous bimanual household manipulation~\cite{gbagbe2024bivla}, while Shake-VLA extends bimanual VLA control to liquid pouring and mixing tasks~\cite{khan2025shakevla}. HapticVLA addresses contact-rich manipulation by transferring tactile-aware behavior into a vision based policy that does not require tactile sensing at inference time~\cite{gubernatorov2026hapticvla}.

Recent work improves spatial grounding in VLA policies through explicit visual cues. VP-VLA {\cite{wang2026vpvlavisualpromptinginterface}} overlays planner-generated target and goal prompts, PointVLA {\cite{li2025pointvlainjecting3dworld}} adds visual grounding to resolve ambiguity, and Gaze2Act {\cite{zuo2026gaze2actgazeconditionedvisionlanguageactionpolicies}} converts human gaze into robot-view masks and interaction points.

Visual Intent Anchors enable preference-aware manipulation by encoding user-selected regions directly in the RGB observation without modifying the VLA backbone.

\section{Method}

\subsection{Visual Intent Anchors}

Visual Intent Anchors encode a user’s preferred grasp and placement choices directly in the RGB observation of a VLA policy.

Let \(I_t \in \mathbf{R}^{H \times W \times 3}\) denote the RGB
observation at timestep \(t\). The user specifies two spatial anchors:
a grasp anchor \(a_g\) on the object and a placement anchor \(a_p\) in
the workspace. These anchors are projected into the image plane and
rendered as binary masks
\[
  M_g, M_p \in \{0,1\}^{H \times W},
\]
where \(M_g\) denotes the grasp-anchor mask and \(M_p\) denotes the
placement-anchor mask. In our implementation, each anchor is rendered
as a disk-shaped image-space mask with radius \(r\). The policy input is
the composited RGB observation
\[
  \tilde{I}_t = \mathrm{overlay}(I_t, M_g, M_p),
\]
where \(\mathrm{overlay}(\cdot)\) denotes alpha compositing of the
active anchor masks onto the RGB frame. In the current fixed-camera Unity implementation, the anchors are stored as image-space masks and re-rendered at each control step. Before grasping, both masks are visible; after grasp completion, \(M_g\) is removed and only \(M_p\) remains active during transport and release. The targets are encoded directly in the RGB observation, allowing the method to work with existing image-conditioned VLA models without architectural changes.

\begin{figure*}[t]
    \centering
    \includegraphics[width=\textwidth]{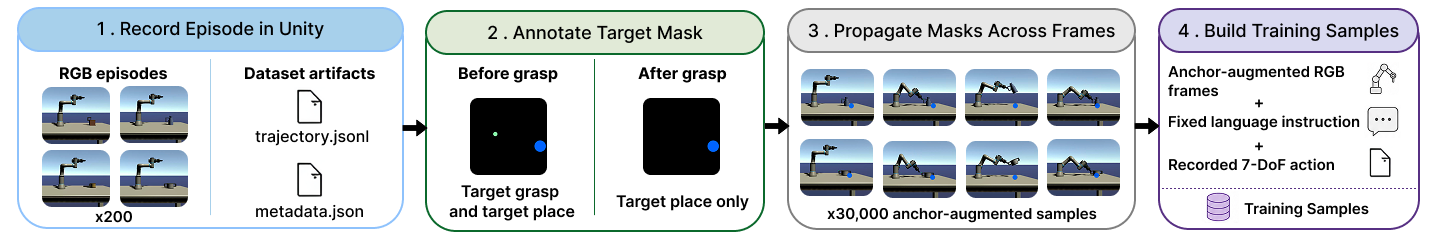}
    \caption{Offline construction of the anchor-augmented dataset. Unity records RGB episodes with trajectories and metadata. Green grasp and blue placement masks specify the user-selected targets and are propagated across the episode, with only the placement mask retained after grasping. Each marked RGB frame is paired with the fixed instruction and corresponding 7-DoF action to form a training sample.}
    \vspace{-.4cm}
    \label{fig:dataset-construction}
\end{figure*}
\subsection{Unity Demonstration Collection}

We implement a Unity environment for demonstration collection, dataset generation, and closed-loop policy execution. It includes a digital twin of a UR3 manipulator, a tabletop workspace, manipulation objects, a fixed camera, and communication with a local OpenVLA server.

The \texttt{DatasetRecorder} stores RGB frames, robot states, end-effector actions, gripper-width commands, task-stage metadata, and annotations. Each episode is saved in an \texttt{episode\_XXXX\_timestamp} directory containing image frames, trajectories, visual-anchor annotations, metadata, and an \texttt{openvla\_train.jsonl} file. Each entry pairs an anchor-augmented image and instruction with the corresponding ground-truth 7-DoF action. The \texttt{BatchEpisodeCollector} increases diversity by varying object poses, target locations, and trajectories. We collect 200 pick-and-place episodes with a mug, pot, thermos, and saucepan, each containing approximately 150 RGB frames.

\subsection{Anchor-Augmented Training Data}

We generate anchor-augmented data offline from recorded Unity demonstrations. For each episode, grasp and placement targets are annotated in the first and final frames and converted into green and blue masks. As shown in Fig.~\ref{fig:dataset-construction}, both masks are propagated until grasp completion, after which only the placement mask remains. Each marked RGB frame is paired with the instruction and corresponding 7-DoF action, then added to the training or validation \texttt{jsonl} file.

\subsection{OpenVLA Fine-Tuning}

The dataset is split at the episode level into 120 training episodes and 80 held-out evaluation episodes, preventing frames from the same trajectory from appearing in both sets. The training split contains 18, 000 anchor-augmented observations.

The anchor-conditioned policy is trained on RGB observations augmented with grasp and placement anchors. During training and inference, it receives the instruction: \textit{``Pick up the object by the green region and place it on the blue region.''} The no-anchor baseline is trained on the corresponding unmodified RGB observations without visual anchors and receives the instruction: \textit{``Pick up the object and place it on the table.''}

Each target action contains translational deltas, controller-defined orientation deltas, and a gripper-width command. Action statistics are computed only on the training split and used for normalization. The normalized actions are converted into target action tokens with the OpenVLA action tokenizer.

We fine-tune OpenVLA-7B with LoRA (r=8, effective batch size 8) to predict tokenized 7-DoF actions from anchor-augmented RGB observations and language instructions. The best-performing LoRA adapter is selected based on validation performance.

\subsection{Online Closed-Loop Unity Execution}

After fine-tuning, the policies are deployed in Unity for closed-loop execution. For the anchor-conditioned policy, the user specifies a grasp target point and a placement target point.


The \texttt{VLAUnityController} captures each RGB frame from the fixed workspace camera, overlays the active anchor masks, and sends the marked image with the language instruction to the local OpenVLA server. The server returns a 7-DoF incremental action containing translational and orientation deltas and a gripper-width command.

Unity executes the action, updates the scene, and renders the next observation, producing a closed-loop rollout conditioned on the outcome of the previous action. The no-anchor baseline uses the same controller, scenes, and execution procedure, but receives unmodified RGB observations.

\section{Experimental Evaluation}
\label{sec} 
We evaluate the policies in closed-loop Unity rollouts on 80 episodes that are disjoint from the 120 training episodes. The quantitative evaluation includes three conditions: (1) \textit{intended anchors}, where the anchor-conditioned policy receives the correct grasp and placement targets; (2) \textit{random object anchors}, where the same policy receives anchors at random locations on the corresponding objects or target surfaces; and (3) a \textit{no-anchor baseline}, trained and evaluated without visual anchors. We also perform an \textit{anchor-following test} with anchors placed at arbitrary workspace locations to determine whether the robot follows their spatial positions.

\subsection{Closed-Loop Evaluation Protocol}

For each held-out episode, all conditions use the same scene, controller, task, and reference demonstration. For the intended-anchor and no-anchor conditions, action-prediction accuracy is evaluated at the grasp and placement events by comparing the predicted translational action components with the corresponding ground-truth actions from the held-out demonstration. The translational error at event \(e\) is computed as:

\begin{equation}
e_e = 100 \sqrt{
\left(\Delta x_e^{\mathrm{p}}-\Delta x_e\right)^2+
\left(\Delta y_e^{\mathrm{p}}-\Delta y_e\right)^2+
\left(\Delta z_e^{\mathrm{p}}-\Delta z_e\right)^2
},
\end{equation}
where \(\Delta x_e^{\mathrm{p}}\), \(\Delta y_e^{\mathrm{p}}\), and
\(\Delta z_e^{\mathrm{p}}\) are the predicted translational action components at event \(e\), while \(\Delta x_e\), \(\Delta y_e\), and \(\Delta z_e\) are the corresponding reference components. The factor of 100 converts the error from meters to centimeters. For conditions with modified anchors, the original demonstration is not used as the target trajectory because a new anchor position implies a different desired action.

All episodes are executed in closed loop without manual intervention or assistive grasping mechanisms. A grasp is considered successful if the robot grasps the object within the specified green region, lifts it by at least 5 cm, and maintains a stable grasp. Placement is considered successful if the robot fully releases the object with its center inside the specified blue region and the object remains stable on the table for at least 1 s. Full-task success requires both stages to be completed within a single rollout before the time limit. For each condition, the success rate is calculated as the number of successful episodes divided by 80. We additionally measure the Euclidean distance between the executed grasp point and the center of the specified grasp region, as well as between the object center after release and the center of the placement region.

\subsection{Results}

The policy successfully completed 73 of 80 held-out rollouts with correct anchors, yielding a full-task success rate of 91.25\%. The mean grasp and placement errors were 0.5~cm and 0.7~cm, respectively, resulting in a mean event-level action error of 0.6~cm. With random anchors on the objects, the mean error increased to 3.0~cm and the success rate decreased to 77.5\% (62 of 80 rollouts), as shown in Fig.~\ref{fig:mean-error}. Thus, correct anchors reduced the average grasp-and-placement action error by a factor of five.
\begin{figure}[!ht]
    \centering
    \includegraphics[width=\columnwidth]{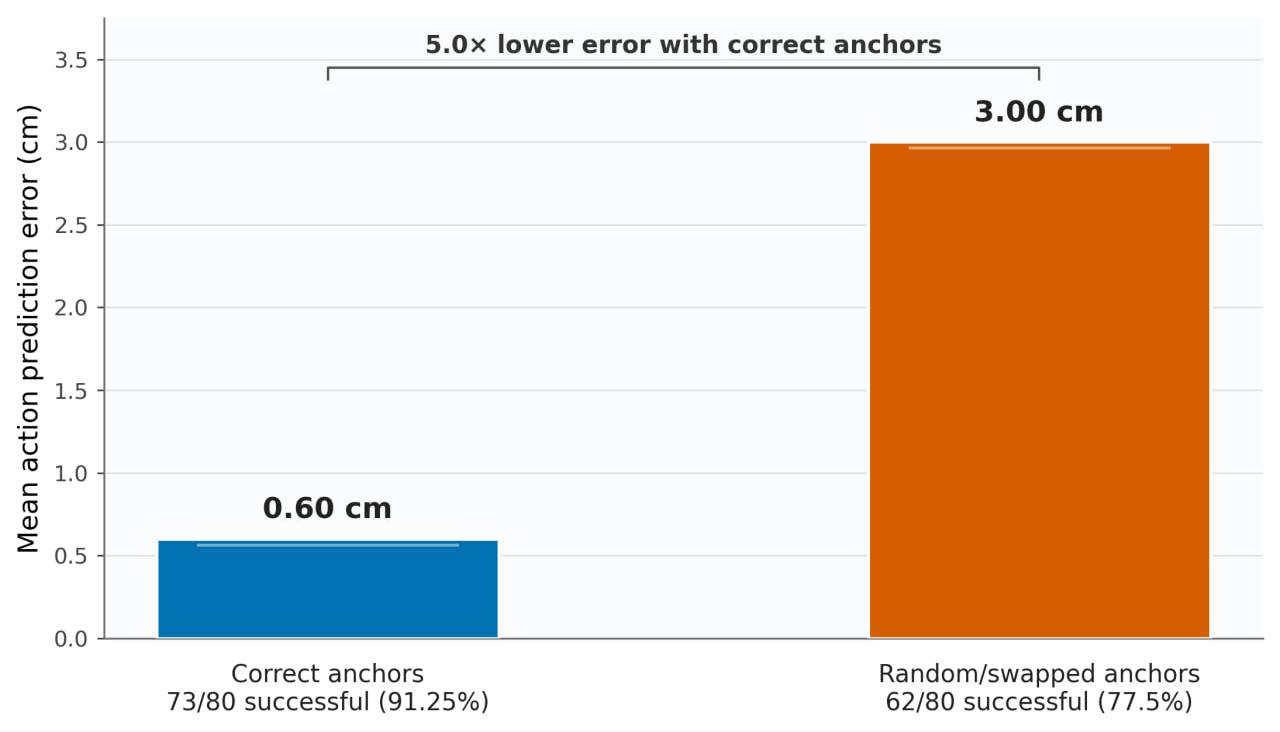}
    \vspace{-5mm}    
    \caption{Mean event-level action error averaged over the grasp and placement events. Correct anchors reduce the error from 3.0~cm to 0.6~cm compared with randomly positioned anchors on the objects.}
    \label{fig:mean-error}
    \vspace{-5mm}
    
\end{figure}

\begin{table*}[!h]
\centering
\caption{Ablation study of visual-anchor conditioning in closed-loop Unity evaluation.}
\label{tab:anchor-ablation}
\setlength{\tabcolsep}{6pt}
\renewcommand{\arraystretch}{1.15}
\begin{tabular}{lccccc}
\toprule
\textbf{Condition} &
\textbf{\shortstack{Grasp\\success}} &
\textbf{\shortstack{Placement\\success}} &
\textbf{\shortstack{Full-task\\success}} &
\textbf{\shortstack{Mean distance to\\grasp anchor center}} &
\textbf{\shortstack{Mean distance to\\placement anchor center}} \\
\midrule

No anchors
& 66.25\%
& 50.0\%
& 50.0\%
& --
& -- \\

Correct anchors
& 95.0\%
& 91.25\%
& 91.25\%
& 0.5 cm
& 0.7 cm \\

Random/swapped anchors
& 85.00\%
& 77.5\%
& 77.5\%
& 2.6 cm
& 3.4 cm \\

\bottomrule
\end{tabular}
\vspace{-5mm}
\end{table*}
The no-anchor baseline achieved 53 of 80 successful grasps (66.25\%) and 40 of 80 successful placements and full-task rollouts (50.0\%). Distance-to-anchor metrics are not reported for this condition because no grasp or placement anchors are provided to the policy. 

Results are summarized in Table~\ref{tab:anchor-ablation}.  Overall, the comparison shows that visual anchors improve closed-loop grasp and placement performance, while correctly positioned anchors produce the highest success rates and the smallest distances to the specified target regions.

In a separate anchor-following test, anchors were placed at arbitrary workspace locations, including positions away from any object. The robot redirected its gripper toward the specified locations rather than toward the objects, showing that the policy actively follows the visual spatial anchors.

\section{Conclusion and Future Work}

This work introduced Visual Intent Anchors, a framework that allows users to specify preferred grasp and placement regions directly in the RGB observation of a vision-language-action policy, reducing ambiguity in spatial intent.

We implemented the method in Unity and fine-tuned OpenVLA-7B with LoRA using anchor-augmented demonstrations across objects with different grasp affordances. With correct anchors, the policy achieved a full-task success rate of 91.25\%, with mean grasp and placement distances of 0.5~cm and 0.7~cm. With anchors randomly positioned on the objects, success decreased to 77.5\%, while the distances increased to 2.6~cm and 3.4~cm. Without anchors, the full-task success rate dropped to 50.0\%; distance-to-anchor metrics are not defined for this condition. In a separate test, anchors placed away from any object redirected the robot toward the specified locations rather than toward the objects. Together, these results show that the policy uses the anchors for spatial conditioning rather than relying only on the scene and language instruction.

From the user perspective, the interaction consists of two direct actions: selecting the grasp and placement targets. This avoids teleoperation, kinematic knowledge, and full trajectory specification. Since the targets are encoded directly in the RGB input, the method remains compatible with OpenVLA and provides a practical basis for VR/AR applications, and human–robot interfaces.

In the current implementation, however, the anchors are stored as fixed image-space masks under a static camera. Consequently, they do not remain attached to an object or workspace location when the camera viewpoint changes or when the object moves. Future work will represent the current image space masks as 3D targets attached to objects or locations in the world coordinate frame.

First, we will convert the existing Unity scene into a VR application using OpenXR and the XR Interaction Toolkit. A user will be able to move around the virtual workspace and select grasp and placement targets directly on object and environment surfaces. In this setting, the Unity digital twin will execute the manipulation. The selected targets will be stored as object-relative or world-space 3D anchors and projected into the simulated robot-camera image at each VLA control step.

Second, we will develop an AR interface for controlling a physical robot. Users will select grasp and placement targets directly on physical objects and workspace surfaces through an AR headset. The robot will provide aligned RGB-D observations, allowing each selected RGB location to be paired with the corresponding depth value and reconstructed as a 3D target in the robot-camera frame. Calibration between the headset, robot base, and RGB-D camera will align these targets with the robot observation, after which they will be reprojected as visual anchors for the VLA policy. Representing the targets as object-relative or world-space 3D anchors, together with continuous headset and object tracking, will preserve their alignment under viewpoint changes, camera motion, and object movement. The policy will predict actions from the marked observation, and the physical robot will execute the manipulation.

 \section*{Acknowledgements} 
Research reported in this publication was financially supported by the RSF grant No. 24-41-02039.


\bibliographystyle{abbrv-doi-etal}
\bibliography{template}
\end{document}